\documentclass[dvips]{arxstspdf}
\usepackage{graphicx}
\usepackage{flushend}
\usepackage{stfloats}

\volume{25}
\issue{4}
\pubyear{2010}
\firstpage{476}
\lastpage{491}
\doi{10.1214/09-STS312}


\begin{document}
\begin{frontmatter}

\title{The EM Algorithm and the Rise of Computational Biology}
\runtitle{EM in Computational Biology}

\begin{aug}
\author[a]{\fnms{Xiaodan} \snm{Fan}\ead[label=e1]{xfan@sta.cuhk.edu.hk}},
\author[b]{\fnms{Yuan} \snm{Yuan}\ead[label=e2]{yuany@google.com}} \and
\author[c]{\fnms{Jun S.} \snm{Liu}\corref{}\ead[label=e3]{jliu@stat.harvard.edu}}
\runauthor{X. Fan, Y. Yuan and J. S. Liu}

\affiliation{The Chinese University of Hong Kong, Google and Harvard University}

\address[a]{Xiaodan Fan is Assistant Professor in Statistics,
Department of Statistics, the Chinese University of Hong Kong, Hong Kong, China
\printead{e1}.}
\address[b]{Yuan Yuan is Quantitative Analyst,
Google, Mountain View, California, USA
\printead{e2}.}
\address[c]{Jun S. Liu is Professor of Statistics,
Department of Statistics, Harvard University, 1 Oxford Street, Cambridge,
Massachusetts 02138, USA
\printead{e3}.}

\end{aug}

%
\begin{abstract}
In the past decade computational biology has grown from a cottage
industry with a handful of researchers to an attractive
interdisciplinary field, catching the attention and imagination of
many quantitatively-minded scientists. Of interest to us is the
key role played by the EM algorithm during this transformation. We
survey the use of the EM algorithm in a few important
computational biology problems surrounding the ``central dogma''
of molecular biology: from DNA to RNA and then to proteins.
Topics of this
article include sequence motif discovery, protein sequence
alignment, population genetics, evolutionary models and mRNA
expression microarray data analysis.
\end{abstract}

%
\begin{keyword}
\kwd{EM algorithm}
\kwd{computational biology}
\kwd{literature review}.
\end{keyword}

\end{frontmatter}
%

\section{Introduction}\label{sec1}

\subsection{Computational Biology}\label{sec1.1}
Started by a few quantitatively minded biologists and biologically
minded mathematicians in the 1970s, computational biology has been
transformed in the past decades to an attractive interdisciplinary
field drawing in many scientists. The use of formal statistical
modeling and computational tools, the expecta\-tion--maximization
(EM) algorithm, in particular, contrib\-uted significantly to this
dramatic transition in solving several key computational biology
problems. Our goal here is to review some of the historical
developments with technical details, illustrating how biology,
traditionally regarded as an empirical science, has come to
embrace rigorous statistical modeling and mathematical reasoning.

Before getting into details of various applications of the EM algorithm
in computational biology, we first explain some basic concepts
of molecular biology. Three kinds of chain biopolymers are the central
molecular building blocks of life: DNA, RNA and proteins. The DNA
molecule is a double-stranded long sequence composed of four types of
nucleotides (A, C, G and T). It has the famous double-helix
structure, and stores the hereditary information. RNA molecules are
very similar to DNAs, composed also of four nucleotides (A, C, G and
U). Proteins are chains of 20 different basic units, called amino
acids.

The genome of an organism generally refers to the collection of
all its DNA molecules, called the chromosomes. Each chromosome
contains both the protein (or RNA) coding regions, called genes,
and noncoding regions. The percentage of the coding regions
varies a lot among genomes of different species. For example, the
coding regions of the genome of baker's yeast are more
than 50\%, whereas those of the human genome are less than 3\%.

RNAs are classified into many types, and the three most basic types
are as follows: messenger RNA (mRNA), transfer
RNA (tRNA) and ribosomal RNA (rRNA). An mRNA can be viewed as an
intermediate copy of its corresponding gene and is used as a template for
constructing the target protein.
tRNA is needed to recruit various amino acids and transport
them to the template mRNA. mRNA, tRNA and amino acids work together
with the construction machineries called ribosomes to make the final
product, protein. One of the main components of ribosomes is the third
kind of RNA, rRNA.

Proteins carry out almost all essential functions in a cell, such
as catalysation, signal transduction, gene regulation, molecular
modification, etc. These capabilities of the protein molecules are
dependent of their 3-dimensional shapes, which, to a
large extent, are uniquely determined by their one-dimensional sequence
compositions. In order to make a protein, the corresponding gene
has to be \textit{transcribed} into mRNA, and then the mRNA is \textit{translated}
into the protein. The ``central dogma'' refers to the
concerted effort of transcription and translation of the cell. The
\textit{expression level} of a gene refers to the amount of its mRNA
in the cell.

Differences between two living organisms are most\-ly due to the
differences in their genomes. Within a~multicellular organism,
however, different cells may differ greatly in both physiology and
function even though they all carry identical genomic information.
These differences are the result of differential gene expression.
Since the mid-1990s, scientists have developed microarray
techniques that can monitor simultaneously the expression levels
of all the genes in a cell, making it possible to construct
the molecular ``signature'' of different cell types. These techniques can
be used to study how a
cell responds to different interventions, and to decipher gene
regulatory networks. A more detailed introduction of the basic
biology for statisticians is given by
\citet{refJWong06Biometrics}.

With the help of the recent biotechnology revolution, biologists
have generated an enormous amount of molecular data, such as
billions of base pairs of DNA sequence data in the GenBank,
protein structure data in PDB, gene expression data, biological
pathway data, biopolymer interaction data, etc. The explosive
growth of various system-level molecu\-lar data calls for
sophisticated statistical models for information integration and for
efficient computational algorithms. Meanwhile,
statisticians have acquired a diverse array of tools for developing such
models and algorithms,
such as the EM algorithm (\citet{refDempsterRubin77JRSSB}), data
augmentation (\citet{reTannerWong87JASA}),
Gibbs sampling
(\citet{refGemanGeman84IEEETransPAMI}),
the Metropolis--Hastings
algorithm (\citet{refMetropolisUlam49JASA};
\citet{refMetropolisTeller53JCP};
\citet{refHastings70Biometrika}), etc.

\subsection{The Expectation--Maximization Algorithm}
The expectation--maximization (EM) algorithm\break
(\citeauthor{refDempsterRubin77JRSSB}, \citeyear{refDempsterRubin77JRSSB}) is
an iterative method for
finding the mode of a marginal likelihood function (e.g.,\vadjust{\goodbreak}
the MLE when there is missing data) or a marginal distribution (e.g., the
maximum {a posteriori} estimator).
Let $\mathbf{Y}$ denote the observed data, $\bolds\Theta$
the parameters of interest, and $\bolds\Gamma$ the nuisance
parameters or missing data. The goal is to
maximize the function
\[
p ( \mathbf{Y} \vert \bolds\Theta) = \int
p( \mathbf{Y}, \bolds\Gamma\vert \bolds\Theta
) \,d \bolds\Gamma,
\]
which cannot be solved
analytically. A basic assumption underlying the effectiveness of
the EM algorithm is that the complete-data likelihood or the posterior
distribution, $p( \mathbf{Y}, \bolds\Gamma
\vert \bolds\Theta)$, is easy to deal with. Starting
with a crude parameter estimate $\bolds\Theta^{(0)}$, the
algorithm iterates between the following Expectation (E-step) and
Maximization (M-step) steps until convergence:

\begin{itemize}
\item E-step: Compute the $Q$-function:
\[
Q \bigl( \bolds\Theta\vert \bolds\Theta^{(t)} \bigr)
\equiv
E_{ \bolds\Gamma\mid\bolds\Theta^{(t)},
\mathbf{Y} } [
\log p (
\mathbf{Y}, \bolds\Gamma\vert \bolds\Theta
)
].
\]
\item M-step: Finding the maximand:
\[
\bolds\Theta^{(t + 1)}
=
\mathop{\operatorname{arg\,max}}_{\bolds\Theta} Q \bigl( \bolds\Theta\vert
\bolds\Theta^{(t)} \bigr).
\]
\end{itemize}

Unlike the Newton--Raphson and scoring algorithms, the
EM algorithm does not require computing the second derivative or
the Hessian matrix. The EM algorithm also has the nice properties of
monotone nondecreasing in the marginal likelihood and stable
convergence to a local mode (or a saddle point) under weak conditions. More
importantly, the EM algorithm is constructed based on the missing
data formulation and often conveys useful statistical insights
regarding the underlying statistical model.
A major drawback of the EM algorithm is that its convergence rate
is only linear, proportional to the fraction of ``missing
information'' about $\bolds\Theta$
(\citet{refDempsterRubin77JRSSB}).
In cases with a large
proportion of missing information, the convergence rate of the EM
algorithm can be very slow. To monitor the convergence rate and
the local mode problem, a basic strategy is to start the EM
algorithm with multiple initial values. More sophisticated methods
are available for specific problems, such as the
``backup-buffering'' strategy in \citet{refQinLiu02AJHG}.

\subsection{Uses of the EM Algorithm in Biology}\label{sec1.3}

The idea of iterating between filling in the missing data and
estimating unknown parameters is so intuitive that some special
forms of the EM algorithm appeared in the literature long before
\citet{refDempsterRubin77JRSSB} defined it. The earliest example
on record is by \citet{refMcKendrick26PEMS}, who invented a
special EM algorithm for fitting a Poisson model to a cholera
infection data set. Other early forms of the EM algorithm
appeared in numerous genetics studies involving allele\vadjust{\goodbreak}
frequency estimation, segregation analysis and pedigree data
analysis (\citeauthor{refCeppelliniSmith55AHG}, \citeyear{refCeppelliniSmith55AHG};
\citeauthor{refSmith57AHG}, \citeyear{refSmith57AHG};
\citeauthor{refOtt79AJHG}, \citeyear{refOtt79AJHG}).
A precursor to the broad recognition of the EM algorithm by the
computational\break biology community is \citet{refChurchill89BMB},
who applied the EM algorithm to fit a hidden Markov model (HMM)
for partitioning genomic sequences into regions with homogenous
base compositions. \citet{refLawrenceReilly90Proteins}
first introduced the EM algorithm for biological sequence motif
discovery. \citet{reHaussleSjolander93ICSS} and
\citet{refKroghHaussler94JMB} formulated an innovative HMM and
used the EM algorithm for protein sequence alignment.
\citet{refKroghHaussler94NAR} extended these
algorithms to predict genes in \textit{E. coli} DNA data.
During the past two decades, probabilistic modeling and the EM
algorithm have become a~more and more common practice in computational
biology, ranging from multiple sequence alignment for a single protein
family (\citeauthor{refDoBatzoglou0GenomeRes}, \citeyear{refDoBatzoglou0GenomeRes})
to genome-wide predictions of protein--protein interactions
(\citeauthor{refDengTing02GenomeRes}, \citeyear{refDengTing02GenomeRes}),
and to single-nucleotide
polymorphism (SNP) haplotype estimation
(\citet{refKangLiu04AJHG}).

As noted in \citet{refMengPedlow92PSCS} and
\citet{refMeng97SMMR}, there are too many EM-related papers to
track. This is true even within the field of
computational biology. In this paper we only examine a few key
topics in computational biology and use typical examples to show
how the EM algorithm has paved the road for these studies. The
connection between the EM algorithm and statistical modeling of
complex systems is essential in computational biology. It is our
hope that this brief survey will stimulate further EM applications
and provide insight for the development of new algorithms.

Discrete sequence data and continuous expression data are two of the most
common data types in computational biology. We discuss sequence
data analysis in Sections \ref{sec2}--\ref{sec5}, and gene expression data
analysis in Section \ref{sec6}. A main objective of computational biology research
surrounding the ``central dogma'' is to study how the gene
sequences affect the gene expression. In Section \ref{sec2} we attempt to
find conserved patterns in functionally related gene sequences as
an effort to explain the relationship of their gene expression. In
Section \ref{sec3} we give an EM algorithm for multiple sequence
alignment, where the goal is to establish ``relatedness'' of
different sequences. Based on the alignment of evolutionary
related DNA sequences, another EM algorithm for detecting
potentially expression-related regions is introduced in Section \ref{sec4}.
An alternative way to deduce the relationship between gene
sequence and gene expression is to check the effect of sequence
variation within the population of a species. In Section \ref{sec5} we
provide an EM algorithm to deal with this type of small sequence
variation. In Section \ref{sec6} we review the clustering analysis of
microarray gene-expression data, which is important for connecting
the phenotype variation among individuals with the expression
level variation. Finally, in Section \ref{sec7} we discuss trends in
computational biology research.

\section{Sequence Motif Discovery and Gene~Regulation}\label{sec2}

\begin{figure}[b]

\includegraphics{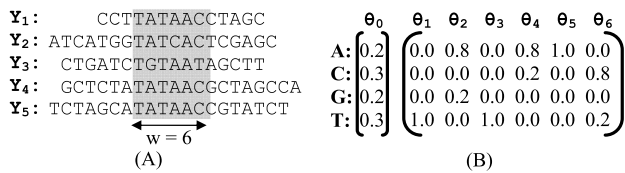}

\caption{Transcription factor binding sites and motifs. \textup{(A)} Each
of the five sequences contains a TFBS
of length 6. The local alignment of these sites is
shown in the gray box. \textup{(B)} The frequency of the
nucleotides outside of the gray box is shown as $\bolds\theta_0$. The frequency
of the nucleotides in the \textit{i}th
column of the gray box is shown as $\bolds\theta_i$. }
\label{fig:motif}
\end{figure}

In order for a gene to be transcribed, special proteins called
transcription factors (TFs) are often required to bind to certain
sequences, called transcription factor binding sites (TFBSs).
These sites are usually 6--20 bp long and are mostly located
upstream of the gene. One TF is usually involved in the regulation
of many genes, and the TFBSs that the TF recognizes often exhibit
strong sequence specificity and conservation (e.g., the first position
of the TFBSs is likely T, etc.). This specific pattern is called a TF
binding motif (TFBM). For example, Figure~\ref{fig:motif} shows a motif
of length 6. The motif is represented by the position-specific
frequency matrix $(\bolds\theta_1,\ldots,\bolds\theta_6)$, which is derived
from the alignment of 5 motif sites by calculating position-dependent frequencies
of the four nucleotides.

In order to understand how genes' mRNA expression levels are
regulated in the cell, it is crucial to identify TFBSs and to
characterize TFBMs. Although much progress has been made in developing
experimental techniques for identifying these TFBSs, these techniques
are typically expensive and time-consuming. They are also limited by
experimental conditions, and cannot pinpoint the binding sites
exactly. In the past twenty years, computational biologists and
statisticians have developed many successful {in silico}
methods to aid biologists in finding TFBSs, and these
efforts have contributed significantly to our understanding of
transcription regulation.\vadjust{\goodbreak}

Likewise, motif discovery for protein sequences is important for
identifying structurally or functionally important regions
(domains) and understanding proteins' functional components, or
active sites. For example, using a Gibbs sampling-based motif
finding algorithm, \citet{refLawrenceWootton93Science} was able to
predict the key \textit{helix-turn-helix}
motif among a family of transcription activators. Experimental approaches
for determining protein motifs are even more expensive and slower
than those for DNAs, whereas computational approaches are more
effective than those for TFBSs predictions.

The underlying logic of computational motif disco\-very is to find
patterns that are ``enriched'' in a~given set of sequence data.
Common methods include word enumeration
(\citet{refSinhaTompa02NAR};
\citet{refHampsonBald02Bioinfo};
\citet{refPavesiPesole04NAR}),
po\-sition-specific frequency matrix
updating (\citet{refStormoHartzell89PNAS};
\citet{refLawrenceReilly90Proteins};
\citet{refLawrenceWootton93Science})
or a combination of the two (\citeauthor{refLiuLiu02NatureBiotech},
\citeyear{refLiuLiu02NatureBiotech}).
The word enumeration approach uses a specific consensus word to
represent a motif. In contrast, the position-specific frequency
matrix approach formulates a motif as a weight matrix.
\citet{refJensenLiu04StatSci} provide a review of these motif
discovery methods. \citet{refTompaZhu05NatureBiotech} compared
the performance of various motif discovery tools. Traditionally,
researchers have employed various heuristics, such as evaluating
excessiveness of word counts or maximizing certain information
criteria to guide motif finding. The EM algorithm was introduced
by \citet{refLawrenceReilly90Proteins} to deal with the motif
finding problem.

As shown in Figure~\ref{fig:motif}, suppose we are given a set of
$K$ sequences $\mathbf{Y} \equiv(\mathbf{Y}_1, \ldots,
\mathbf{Y}_K)$, where $\mathbf{Y}_k \equiv(Y_{k,1}, \ldots,
Y_{k,L_k})$ and $Y_{k,l}$ takes values in an alphabet of $d$
residues ($d=4$ for DNA/RNA and 20 for protein). The alphabet is
denoted by $\mathbf{R} \equiv(r_1,\ldots,r_d)$. Motif sites in
this paper refer to a set of contiguous segments of the same
length $w$ (e.g., the marked 6-mers in Figure~\ref{fig:motif}). This
concept can be further generalized via a hidden Markov model to
allow gaps and position deletions (see
Section~\ref{sec3} for HMM discussions). The weight
matrix, or \textit{Product-Multinomial} motif model, was first
introduced by \citet{refStormoHartzell89PNAS} and later
formulated rigorously in \citet{refLiuLawrence95JASA}. It
assumes that, if $Y_{k,l}$ is the \textit{i}th position of a
motif site, it follows the multinomial distribution with the
probability vector $\bolds\theta_i \equiv(\theta_{i1},
\ldots, \theta_{id})$; we
denote this model as $\mathit{PM}(\bolds\theta_1, \ldots,
\bolds\theta_w)$. If $Y_{k,l}$ does not belong to any
motif site, it is generated\vadjust{\goodbreak} independently from the multinomial
distribution with parameter
$\bolds\theta_0 \equiv(\theta_{01}, \ldots, \theta_{0d})$.

Let $\bolds\Theta\equiv(\bolds\theta_0, \bolds\theta_1, \ldots, \bolds\theta_w)$.
For sequence
$\mathbf{Y}_k$, there are $L'_k=L_k-w+1$ possible positions a
motif site of length $w$ may start.
To represent the motif locations, we introduce
the unobserved indicators $\bolds\Gamma\equiv
\{\Gamma_{k,l} \mid1 \leq k \leq K, 1 \leq l \leq L'_k \}$, where
$\Gamma_{k,l}=1$ if a motif site starts at position $l$ in
sequence $\mathbf{Y}_k$, and $\Gamma_{k,l}=0$ otherwise. As
shown in Figure~\ref{fig:motif}, it is straightforward to estimate
$\bolds\Theta$ if we know where the motif sites are.
The motif location indicators $\bolds\Gamma$ are
the missing data that makes the EM framework a natural choice
for this problem. For illustration, we further assume
that there is exactly one motif site within each sequence and that its
location in the sequence is uniformly distributed. This means that
$\sum_l \Gamma_{k,l} =1$ for all $k$ and $P(\Gamma_{k,l}=1) =
\frac{1}{L'_k}$.

Given $\Gamma_{k,l} =1$, the probability of each observed sequence
$\mathbf{Y}_k$ is
%
\begin{equation}\label{motif-complete}
\small{ P(\mathbf{Y}_k \vert \Gamma_{k,l}=1, \bolds\Theta) =
\bolds\theta^{\mathbf{h}(\mathbf{B}_{k,l})}_0 \prod^{w}_{j=1}
\bolds\theta^{\mathbf{h}(\mathbf{Y}_{k,l+j-1})}_i.}
\end{equation}
In this expression, $\mathbf{B}_{k,l} \equiv\{Y_{k,j}\dvtx  j<l \mbox{
or }
j\geq l+w\}$ is the set of letters of nonsite positions of
$\mathbf{Y}_k$. The counting function $\mathbf{h}(\cdot)$ takes a
set of letter symbols as input and outputs the column vector
$(n_1,\ldots,n_d)^T$, where $n_i$ is the number of base type $r_i$
in the input set. We define the vector power function as
$\bolds\theta^{\mathbf{h}(\cdot)}_i\equiv\prod^d_{j=1}
\theta^{n_j}_{ij}$ for\vspace*{1pt} $i=0,\ldots,w$. Thus, the complete-data
likelihood function is the product of equation
(\ref{motif-complete}) for $k$ from 1 to $K$, that is,\vspace*{-1pt}
%
\begin{eqnarray*}
P(\mathbf{Y}, \bolds\Gamma\vert \bolds\Theta) &\propto& \prod^K_{k=1}
\prod^{L'_k}_{l=1} P(\mathbf{Y}_k
\vert \Gamma_{k,l}=1, \bolds\Theta)^{\Gamma_{k,l}} \\
&= & \bolds\theta^{\mathbf{h}(\mathbf{B}_{\bolds\Gamma})}_0 \prod^{w}_{i=1}
\bolds\theta^{\mathbf{h}(\mathbf{M}^{(i)}_{\bolds\Gamma})}_i,
\end{eqnarray*}
where $\mathbf{B}_{\bolds\Gamma}$ is the set of all
nonsite bases, 
and $\mathbf{M}^{(i)}_{\bolds\Gamma}$ is the set of
nucleotide bases at position $i$ of the TFBSs given the indicators
${\bolds\Gamma}$.

The MLE of $\bolds\Theta$ from the complete-data likelihood
can be determined by simple counting, that is,
\[
\hat{\bolds\theta}_i = \frac{\mathbf{h}(\mathbf{M}^{(i)}_{\bolds\Gamma})}{K}
\quad  \mbox{and}\quad
\hat{\bolds\theta}_0 = \frac{\mathbf{h}(\mathbf{B}_{\bolds\Gamma})}{\sum^K_{k=1}
(L_k - w)}.
\]
The EM algorithm for this problem is quite intuitive. In the
E-step, one uses the current parameter values $\bolds{\Theta}^{(t)}$ to
compute the expected values of $h(\mathbf{M}^{(i)}_{\bolds\Gamma})$ and
$h(\mathbf{B}_{\bolds\Gamma})$.
More precisely, for sequence $Y_k$, we compute its
likelihood of being generated from $\bolds{\Theta}^{(t)}$ conditional on
each possible motif location $\Gamma_{k,l}=1$,
\begin{eqnarray*}
w_{k,l} &\equiv& P\bigl(\mathbf{Y}_k \vert \Gamma_{k,l}=1,
\bolds\Theta^{(t)}\bigr)\\
& =&
\biggl(\frac{\bolds\theta_{1}}{\bolds\theta_{0}}
\biggr)^{\mathbf{h}(Y_{k,l})} \cdots
\biggl(\frac{\bolds\theta_w}{\bolds\theta_0}
\biggr)^{\mathbf{h}(Y_{k,l+w-1})}\bolds\theta_{0}^{\mathbf
{h}(\mathbf{Y}_k)}.
\end{eqnarray*}
Letting $W_k \equiv\sum_{l=1}^{L'_k} w_{k,l}$, we then
compute the expected count vectors as
\begin{eqnarray*}
E_{ \bolds\Gamma\mid\bolds\Theta^{(t)}, \mathbf{Y} } \bigl[\mathbf{h}
\bigl(\mathbf{M}^{(i)}_{\bolds\Gamma}\bigr)\bigr]
&=&
\sum^K_{k=1} \sum^{L'_k}_{l=1} \frac{w_{k,l}}{W_k}
\mathbf{h}(Y_{k,l+i-1}),\\
E_{ \bolds\Gamma\mid\bolds\Theta^{(t)}, \mathbf{Y} } [\mathbf{h}
(\mathbf{B}_{\bolds\Gamma})]
&=&
\mathbf{h}(\{Y_{k,l} \dvtx  1 \le k \le K, 1 \le l \le L_k\})\\
&&{}- \sum^w_{i=1} E_{ \bolds\Gamma\mid\bolds\Theta^{(t)}, \mathbf{Y} }
\bigl[\mathbf{h}\bigl(\mathbf{M}^{(i)}_{\bolds\Gamma}\bigr)\bigr].
\end{eqnarray*}
In the M-step, one simply computes
\begin{eqnarray*}
\bolds\theta^{(t+1)}_i &=& \frac{E_{
\bolds\Gamma\mid\bolds\Theta^{(t)}, \mathbf{Y} }
[\mathbf{h}(\mathbf{M}^{(i)}_{\bolds\Gamma})]}{K}
\quad  \mbox{and}\\
\bolds\theta^{(t+1)}_0 &=& \frac{E_{ \bolds\Gamma
\mid\bolds\Theta^{(t)}, \mathbf{Y} } [\mathbf{h}(\mathbf{B}_{\bolds\Gamma})]}
{\sum^K_{k=1} (L_k - w)} .
\end{eqnarray*}
It is necessary to start with a nonzero initial weight matrix
$\bolds\Theta^{(0)}$ so as to guarantee that $P(\mathbf{Y}_k\vert \Gamma_{k,l}=1,
\bolds\Theta^{(t)}) >0$ for all
$l$. At convergence the algorithm yields both the MLE
$\hat{\bolds\Theta}$ and predictive probabilities for candidate TFBS locations,
that is, $P(\Gamma_{k,l}=1 \vert \hat{\bolds\Theta},\mathbf{Y})$.

\citet{refCardonStormo92JMB} generalized the above simple
model to accommodate insertions of variable lengths in the middle
of a binding site. To overcome the restriction that each sequence
contains exactly one motif site, \citeauthor{refBaileyElkan94ISMB}
(\citeyear{refBaileyElkan94ISMB},
\citeyear{refBaileyElkan95MachineLearning},
\citeyear{refBaileyElkan95ISMB})
introduced a parameter $p_0$ describing the prior probability for
each sequence position to be the start of a motif site, and
designed a modified EM algorithm called the \underline{M}ultiple
\underline{E}M for \underline{M}otif \underline{E}licitation (MEME).
Independently, \citet{refLiuLawrence95JASA}
presented a full Bayesian framework and Gibbs sampling algorithm for
this problem. Compared with the EM approach, the Markov chain Monte
Carlo (MCMC)-based approach has the advantages of making more flexible
moves during the iteration and incorporating additional information
such as motif location and orientation preference in the model.

The generalizations in \citet{refBaileyElkan94ISMB} and
\citet{refLiuLawrence95JASA} assume that all overlapping
subsequences of length $w$ in the sequence data set are from a
finite mixture model. More precisely, each subsequence of length
$w$ is trea\-ted as~an independent sample from a mixture of
$\mathit{PM}(\bolds\theta_1, \ldots, \bolds\theta_w)$ and
$\mathit{PM}(\bolds\theta_0, \ldots, \bolds\theta_0)$
[independent\break $\operatorname{Multinomial}(\theta_0)$ in all $w$ positions]. The EM
solution of this mixture model formulation then leads to the MEME
algorithm of \citet{refBaileyElkan94ISMB}. To deal with the
situation that $w$ may not be known precisely, MEME searches
motifs of a range of different widths separately, and then
performs model selection by optimizing a heuristic function based
on the maximum likelihood ratio test. Since its release, MEME has
been one of the most popular motif discovery tools cited in the
literature. The Google scholar search gives a count of 1397
citations as of August 30th, 2009. Although it is 15 years old,
its performance is still comparable to many new algorithms
(\citeauthor{refTompaZhu05NatureBiotech}, \citeyear{refTompaZhu05NatureBiotech}).

\section{Multiple Sequence Alignment}\label{sec3}

Multiple sequence alignment (MSA) is an important tool for
studying structures, functions and the evolution of proteins. Because
different parts of a protein may have different functions, they
are subject to different selection pressures during evolution.
Regions of greater functional or structural importance are
generally more conserved than other regions. Thus, a good alignment
of protein sequences can yield
important evidence about their functional and structural properties.

Many heuristic methods have been proposed to~sol\-ve the MSA
problem. A popular approach is the~progressive alignment method
(\citeauthor{refFengDoolittle87JME}, \citeyear{refFengDoolittle87JME}),
in which the MSA is built up
by aligning the most closely related sequences first
and then adding more distant sequences successively. Many
alignment programs are based on this strategy, such as MULTALIGN
(\citeauthor{refBartonSternberg87JMB}, \citeyear{refBartonSternberg87JMB}), MULTAL
(\citeauthor{refTaylor88JME}, \citeyear{refTaylor88JME}) and, the most influential
one, ClustalW
(\citeauthor{refThompsonGibson94NAR}, \citeyear{refThompsonGibson94NAR}).
Usually, a \textit{guide tree}
based on pairwise similarities between the protein sequences is
constructed prior to the multiple alignment to determine the order
for sequences to enter the alignment. Recently, a few new progressive
alignment algorithms with significantly improved alignment
accuracies and speed have been proposed, including T-Coffee
(\citet{refNotredameHeringa00JMB}),
MAFFT (\citeauthor{reKatohMiyata05NAR}, \citeyear{reKatohMiyata05NAR}), PROBCONS
(\citeauthor{refDoBatzoglou0GenomeRes}, \citeyear{refDoBatzoglou0GenomeRes}) and
MUSCLE
(\citeauthor{reEdga04NAR}, \citeyear{refEdga04BMCBioinfo}, \citeyear{reEdga04NAR}).
They differ from previous approaches and each other mainly in the construction
of the guide tree and in the objective function for judging the
goodness of the alignment. \citet{refBatzoglou05BriefBioinfo}
and \citet{refWallaceHiggins05CurOpiStructBio} reviewed these
algorithms.

\begin{figure}

\includegraphics{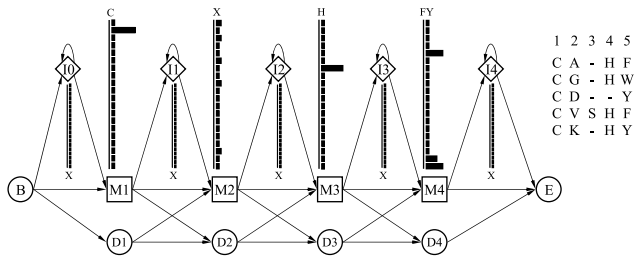}

\caption{Profile hidden Markov model. A modified toy example is
adopted from Eddy \textup{(\protect\citeyear{reEddy98Bioinfo})}.
It shows the alignment
of five sequences, each containing only three to five letters. The
first position is enriched with Cysteine (C), the fourth position
is enriched with Histidine (H), and the fifth position is enriched
with Phenylalanine (F) and Tyrosine (Y). The third sequence has a
deletion at the fourth position, and the fourth sequence has an
insertion at the third position. This simplified model does not
allow insertion and deletion states to follow each other.} \label{fig:profileHMM}
\end{figure}

An important breakthrough in solving the MSA problem is the
introduction of a probabilistic generative model, the profile
hidden Markov model by \citet{refKroghHaussler94JMB}.
The profile HMM postulates that the $N$ observed sequences are generated
as independent but indirect observations (\textit{emissions}) from a
Markov chain model illustrated in Figure~\ref{fig:profileHMM}. The
underlying unobserved Markov chain consists of three types of
states: match, insertion and deletion. Each match or insertion
state \textit{emits} a letter chosen from the alphabet $\mathbf{R}$
(size $d=20$ for proteins) according to a multinomial
distribution. The deletion state does not emit any letter, but
makes the sequence generating process skip one or more
match states. A multiple alignment of the $N$ sequences is
produced by aligning the letters that are emitted from the same
match state.

Let $\bolds{\Gamma}_i$ denote the unobserved state path
through which the \textit{i}th sequence is generated from the
profile HMM, and $\mathbf{S}$ the set of all states. Let
$\bolds\Theta$ denote the set of all global parameters of
this model, including emission probabilities in match and
insertion states $e_{lr}$ $(l \in\mathbf{S}, r \in\mathbf{R})$, and
transition probabilities among all hidden states
$t_{ab}$
$(a,b \in\mathbf{S})$. The complete-data log-likelihood
function can be written as
\begin{eqnarray*}
&&\log P(\mathbf{Y}, \bolds\Gamma|\bolds\Theta)\\
 &&\quad =
\sum_{i=1}^N [\log P(\mathbf{Y}_i|\bolds\Gamma_i,
\bolds\Theta)+\log P(\bolds\Gamma_i|\bolds\Theta) ]\\
&&\quad =  \sum_{i=1}^N \biggl[ \sum_{l \in\mathbf{S}, r \in
\mathbf{R}} {M_{lr}(\bolds\Gamma_i)} \log e_{lr}\\
&&\hspace*{42pt} {} +
\sum_{a,b \in\mathbf{S}} {N_{ab}(\bolds\Gamma_i)}\log
t_{ab} \biggr],
\end{eqnarray*}
where $M_{lr}(\bolds\Gamma_i)$ is the count of letter $r$ in
sequence $\mathbf{Y}_i$ that is generated from state $l$
according to $\bolds\Gamma_i$, and $N_{ab}(\bolds\Gamma_i)$ is the count
of state transitions from $a$ to $b$ in
the path $\bolds\Gamma_i$ for
sequence $\mathbf{Y}_i$.

The E-step involves calculating the expected counts of emissions
and transitions, that is, $E[ M_{lr}(\bolds\Gamma_i)
\vert \bolds{\Theta}^{(t)}]$ and $E[N_{ab}(\bolds\Gamma_i)
\vert \bolds{\Theta}^{(t)}]$,
averaging over all possible generating paths $\bolds\Gamma_i$. The $Q$-function is
\begin{eqnarray*}
Q\bigl(\bolds\Theta|\bolds\Theta^{(t)}\bigr) &=& \sum_{i=1}^N
\sum_{\bolds\Gamma_i} \frac{P(\bolds\Gamma_i,\mathbf{Y}_i|\bolds\Theta^{(t)})}
{P(\mathbf{Y}_i|\bolds\Theta^{(t)})} \\
&&\hspace*{33pt}{}\cdot\biggl[ \sum_{l \in\mathbf{S},r
\in\mathbf{R}} \log(e_{lr}) M_{lr}(\bolds\Gamma_i)\\
&&\hspace*{43pt}{} +
\sum_{a,b \in\mathbf{S}} \log(t_{ab}) N_{ab}(\bolds\Gamma_i) \biggr].
\end{eqnarray*}
A brute-force enumeration of all paths is prohibitively expensive in
computation.
Fortunately, one can apply a forward--backward dynamic
programming technique to compute the expectations for each
sequence and then sum them all up.

In the M-step, the emission and transition probabilities are
updated as the ratio of the expected event occurrences (sufficient
statistics) divided by the total expected emission or transition
events:
\begin{eqnarray*}
e_{lr}^{(t+1)} &=& \frac{ \sum_i \{m_{lr}(\mathbf{Y}_i)/P
(\mathbf{Y}_i|\bolds\Theta^{(t)})\}}{\sum_i
\{m_l(\mathbf{Y}_i)/P(\mathbf{Y}_i| \bolds\Theta^{(t)})\}
}, \\
t_{ab}^{(t+1)} &=& \frac{ \sum_i \{ n_{ab}(\mathbf{Y}_i)/P
(\mathbf{Y}_i| \bolds\Theta^{(t)})\}}{\sum_i
\{n_a(\mathbf{Y}_i)/P(\mathbf{Y}_i|\bolds\Theta^{(t)})\}},
\end{eqnarray*}
where
\begin{eqnarray*}
m_{lr}(\mathbf{Y}_i) &=& \sum_{\bolds\Gamma_i}
M_{lr}(\bolds\Gamma_i)P\bigl(\bolds\Gamma_i,\mathbf{Y}_i|\bolds\Theta^{(t)}\bigr),\\
n_{ab}(\mathbf{Y}_i) &=& \sum_{\bolds\Gamma_i}
N_{ab}(\bolds\Gamma_i)P\bigl(\bolds\Gamma_i,\mathbf{Y}_i|\bolds\Theta^{(t)}\bigr),\\
m_l(\mathbf{Y}_i) &=& \sum_{r \in\mathbf{R}}
m_{lr}(\mathbf{Y}_i),\quad    n_a(\mathbf{Y}_i) = \sum_{b \in
\mathbf{S}} n_{ab}(\mathbf{Y}_i).
\end{eqnarray*}
%
This method is called the Baum--Welch algorithm
(\citeauthor{reBaumWeiss70AMS}, \citeyear{reBaumWeiss70AMS}), and is
mathematically equivalent to
the EM algorithm. Conditional on the MLE $\hat{\bolds\Theta}$, the best
alignment path for each sequence can be found
efficiently by the Viterbi algorithm (see
\citeauthor{refDurbinMitchison98book}, \citeyear{refDurbinMitchison98book},
Chapter 5, for details).

The profile HMM provides a rigorous statistical modeling and inference
framework for the MSA problem. It has also played a
central role in advancing the understanding of protein families and domains.
A protein family database, Pfam
(\citeauthor{refFinnBateman06NAR}, \citeyear{refFinnBateman06NAR}), has been
built using profile HMM
and has served as an essential source of data in the
field of protein structure and function research.
Currently there are two popular software packages that use
profile HMMs to detect remote protein homologies:
HMMER (\citeauthor{reEddy98Bioinfo}, \citeyear{reEddy98Bioinfo}) and SAM
(\citeauthor{refHugheyKrogh96CompApplBio}, \citeyear{refHugheyKrogh96CompApplBio};
\citeauthor{reKarplusHughey99Bioinfo}, \citeyear{reKarplusHughey99Bioinfo}).
\citet{refMaderaGough02NAR} gave a comparison of
these two packages.

There are several challenges in fitting the profile HMM. First, the size
of the model (the number of match, insertion and deletion states)
needs to be determined before model fitting. It is common to begin
fitting a profile HMM by setting the number of match states equal to the
average sequence length. Afterward, a strategy called ``model
surgery'' (\citeauthor{refKroghHaussler94JMB}, \citeyear{refKroghHaussler94JMB})
can be applied to
adjust the model size (by adding or removing a match state
depending on whether an insertion or a deletion is used too
often). \citet{reEddy98Bioinfo} used a \textit{maximum a
posteriori} (MAP) strategy to determine the model size in HMMER.
In this method the number of match states is given a prior distribution,
which is equivalent to adding a penalty term in the log-likelihood
function.

Second, the number of sequences is sometimes too small for
parameter estimation. When calculating the conditional
expectation of the sufficient statistics, which are counts of
residues at each state and state transitions, there may not be
enough data, resulting in zero counts which could make the estimation
unstable. To avoid the occurrence of zero counts,
pseudo-counts can be added. This is equivalent to using a
Dirichlet prior for the multinomial parameters in a Bayesian
formulation.

Third, the assumption of sequence independence is often violated.
Due to the underlying evolutionary relationship (unknown), some of
the sequences may share much higher mutual similarities than
others. Therefore, treating all sequences as i.i.d. samples may
cause serious biases in parameter estimation. One possible
solution is to give each sequence a weight according to its
importance. For example, if two sequences are identical, it is
reasonable to give each of them half the weight of other
sequences. The weights can be easily integrated into the M-step of
the EM algorithm to update the model parameters. For example, when
a sequence has a weight of 0.5, all the emission and transition
events contributed by this sequence will be counted by half.
Many methods have been proposed to assign weights to the sequences
(\citeauthor{refDurbinMitchison98book}, \citeyear{refDurbinMitchison98book}),
but it is not clear how to
set the weights in a principled way to best account for sequence
dependency.

Last, since the EM algorithm can only find local modes of the
likelihood function, some stochastic perturbation can be
introduced to help find better modes and improve the alignment.
Starting from multiple random initial parameters is strongly
recommended. \citet{refKroghHaussler94JMB} combined simulated
annealing into Baum--Welch and showed some improvement.
\citet{refBaldiChauvin94NeuralComp} developed a generalized EM
(GEM) algorithm using a gradient ascent calculation in an attempt
to infer HMM parameters in a smoother way.

Despite many advantages of the profile HMM, it is no longer the
mainstream MSA tool. A main reason is that the model has too many
free parameters, which render the parameter estimation very
unstable when there are not enough sequences (fewer than 50, say)
in the alignment. In addition, the vanilla EM algorithm and its
variations developed by early researchers for the MSA problem
almost always converge to suboptimal alignments. Recently,
\citet{refEdlefsenLiu0Unknown} have developed an ECM
algorithm for MSA that appears to have much improved convergence
properties. It is also difficult for the profile HMM to
incorporate other kinds of information, such as 3D protein
structure and guide tree. Some recent programs such as 3D-Coffee
(\citeauthor{reOSullivanNotredame04JMB}, \citeyear{reOSullivanNotredame04JMB})
and MAFFT are more flexible
as they can incorporate this information into the objective
function and optimize it. We believe that the Monte Carlo-based
Bayesian approaches, which can impose more model constraints
(e.g., to capitalize on the ``motif'' concept) and make more
flexible MCMC moves, might be a promising route to rescue profile
HMM (see \citeauthor{refLiuLawrence95JASA}, \citeyear{refLiuLawrence95JASA};
\citeauthor{refNeuwaldLiu04BMCBioinfo}, \citeyear{refNeuwaldLiu04BMCBioinfo}).

\section{Comparative Genomics}\label{sec4}

A main goal of comparative genomics is to identify and
characterize functionally important regions in the genome of
multiple species. An assumption underlying such studies is that,
due to evolutionary pressure, functional regions in the genome
evolve much more slowly than most nonfunctional regions due to
functional constraints (\citeauthor{refWolfeLi89Nature},
\citeyear{refWolfeLi89Nature};
\citeauthor{refBoffelliRubin03Science}, \citeyear{refBoffelliRubin03Science}).
Regions that evolve more slowly
than the background are called evolutionarily conserved elements.

Conservation analysis (comparing genomes of related species) is a
powerful tool for identifying functional elements such as
protein/RNA coding regions and transcriptional regulatory
elements. It begins with an alignment of multiple orthologous
sequences (sequences evolved from the same common ancestral
sequence) and a conservation score for each column of the
alignment. The scores are calculated based on the likelihood that
each column is located in a conserved element. The phylogenetic
hidden Markov model (Phylo-HMM) was introduced to infer the
conserved regions in the genome (\citeauthor{refYang95Genetics},
\citeyear{refYang95Genetics};
\citeauthor{refFelsensteinChurchill96MolBiolEvol},
\citeyear{refFelsensteinChurchill96MolBiolEvol};
\citeauthor{refSiepeHaussler05GenomeRes}, \citeyear{refSiepeHaussler05GenomeRes}).
The statistical power of
Phylo-HMM has been systematically studied by
\citet{refFanLiu07BMCBioinfo}.
\citet{refSiepeHaussler05GenomeRes} used the EM algorithm for
estimating parameters in Phylo-HMM. Their results, provided by the
UCSC genome browser database (\citeauthor{refKarolchikKent03NAR},
\citeyear{refKarolchikKent03NAR}),
are very influential in the computational biology community. By
August~2009, the paper of \citet{refSiepeHaussler05GenomeRes}
had been cited 413 times according to the Web of Science database.

\begin{figure}

\includegraphics{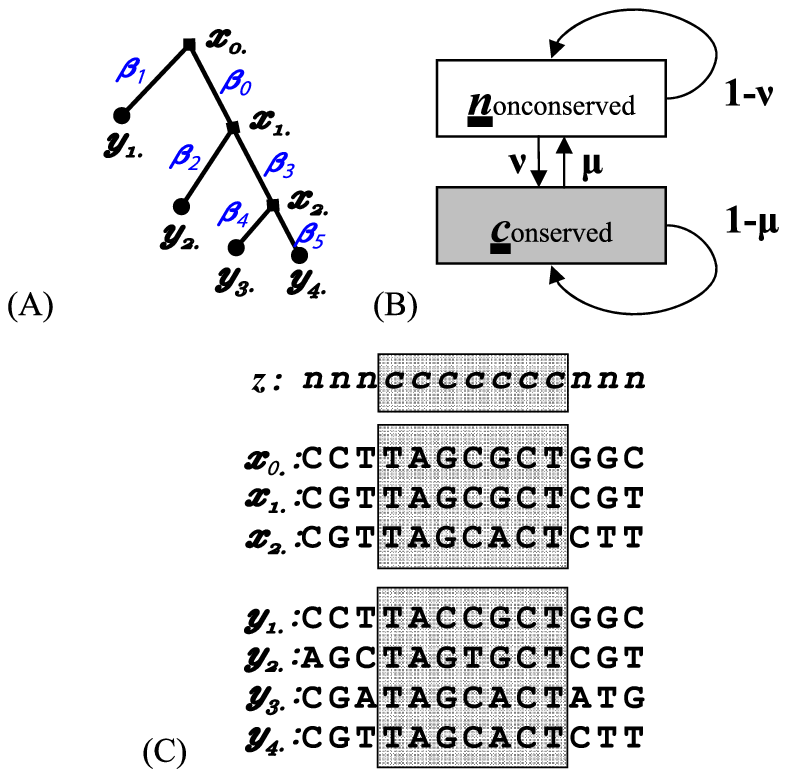}

\caption{Two-state Phylo-HMM. \textup{(A)} Phylogenetic tree: The tree
shows the evolutionary relationship of four contemporary sequences
($\mathbf{y}_{1\cdot}, \mathbf{y}_{2\cdot}, \mathbf{y}_{3\cdot},
\mathbf{y}_{4\cdot}$).
They are evolved from the
common ancestral sequence $\mathbf{x}_{0\cdot}$, with two
additional internal nodes (ancestors), $\mathbf{x}_{1\cdot}$
and $\mathbf{x}_{2\cdot}$. The branch lengths $\bolds\beta=
(\beta_0,\beta_1,\beta_2,\beta_3,\beta_4,\beta_5)$
indicate the evolutionary distance between two nodes, which are
measured by the expected number of substitutions per site. \textup{(B)}~HMM
state-transition diagram: The system consists of a state for
conserved sites and a state for nonconserved sites (c and n,
respectively). The two states are associated with different
phylogenetic models ($\psi_c$ and $\psi_n$), which differ by a
scaling parameter $\rho$. \textup{(C)} An illustrative alignment generated
by this model: A~state sequence ($z$) is generated according to
$\mu$ and $\nu$. For each site in the state sequence, a nucleotide
is generated for the root node in the phylogenetic tree and then
for subsequent child nodes according to the phylogenetic model
($\psi_c$ or $\psi_n$). The observed alignment $\mathbf{Y} =
(\mathbf{y}_{1\cdot}, \mathbf{y}_{2\cdot}, \mathbf{y}_{3\cdot},
\mathbf{y}_{4\cdot})$
is composed of all
nucleotides in the leaf nodes. The state sequence $\mathbf{z}$
and all ancestral sequences $\mathbf{X} = (\mathbf{x}_{0\cdot},
\mathbf{x}_{1\cdot},
\mathbf{x}_{2\cdot})$ are
unobserved.} \label{fig:phyloHMM} 
\end{figure}

As shown in Figure~\ref{fig:phyloHMM}, the alignment modeled by
Phylo-HMM can be seen as generated from two steps. First, a
sequence of $L$ sites is generated from a two-state HMM, with the
hidden states being conserved or nonconserved sites. Second, a
nucleotide is generated for each site of the common ancestral
sequence and evolved to the contemporary nucleotides along all
branches of a phylogenetic tree independently according to the
corresponding phylogenetic model.

Let $\mu$ and $\nu$ be the transition probabilities between the
two states, and let the phylogenetic models for
nonconserved and conserved states be
$\psi_n = (Q, \pi, \tau, \bolds\beta)$ and
$\psi_c = (Q, \pi, \tau, \rho\bolds\beta)$, respectively.
Here $\pi$ is the emission probability vector of the four
nucleotides (A, C, G and T) in the common ancestral sequence
$\mathbf{x}_0$; $\tau$~is the tree topology of the
corresponding phylogeny; $\bolds\beta$ is a vector of
non-negative real numbers representing branch lengths of the tree,
which are measured by the expected number of substitutions per
site. The difference between the two states is characterized by a
scaling parameter $\rho\in[0,1)$ applied to the branch lengths
of only the conserved state, which means fewer substitutions. The
nucleotide substitution model considers a descendent nucleotide to
have evolved from its ancestor by a continuous-time
time-homogeneous Markov process with transition kernel $Q$, also
called the substitution rate matrix
(\citeauthor{refTavare86Lectures}, \citeyear{refTavare86Lectures}).
The transition kernels for all
branches are assumed to be the same. Many parametric forms are
available for the \mbox{4-by-4} nucleotide substitution rate matrix $Q$,
such as the Jukes--Cantor substitution matrix and the general
time-reversible substitution matrix
(\citeauthor{refYang97ComputApplBiosci}, \citeyear{refYang97ComputApplBiosci}).
The nucleotide
transition probability matrix for a branch of length $\beta_i$ is
$e ^ {\beta_i Q}$.

\citet{refSiepeHaussler05GenomeRes} assumed that the tree
topology $\tau$ and the emission probability vector $\pi$ are
known. In this case, the observed alignment $\mathbf{Y} =
(\mathbf{y}_{1\cdot}, \mathbf{y}_{2\cdot}, \mathbf{y}_{3\cdot},
\mathbf{y}_{4\cdot})$
is a matrix of nucleotides.
The parameter of interest is $\bolds\Theta= (\mu, \nu, Q,
\rho, \bolds\beta)$. The missing information $\bolds\Gamma= (\mathbf{z},
\mathbf{X})$
includes the state sequence $\mathbf{z}$ and the ancestral
DNA sequences $\mathbf{X}$. The complete-data likelihood is
written as
\begin{eqnarray*}
&&P(\mathbf{Y}, \bolds\Gamma| \bolds\Theta) \\
&&\quad =
b_{z_{1}} P(\mathbf{y}_{\cdot1}, \mathbf{x}_{\cdot1} |
\psi_{z_1}) \prod^{L}_{i=2} a_{z_{i-1} z_i} P(\mathbf{y}_{\cdot
i}, \mathbf{x}_{\cdot i}| \psi_{z_i}).
\end{eqnarray*}
Here $\mathbf{y}_{\cdot i}$ is the \textit{i}th column of the
alignment $\mathbf{Y}$, $z_i \in\{c,n\}$ is the hidden state
of the \textit{i}th column, $(b_c, b_n)=(\frac{\nu}{\mu+ \nu},
\frac{\mu}{\mu+ \nu})$ is the initial state probability of the
HMM\vspace*{1.5pt} if the chain is stationary, and $a_{z_{i-1} z_i}$ is the
transition probability (as illustrated in Figure~\ref{fig:phyloHMM}).

The EM algorithm is applied to obtain the MLE of $\bolds\Theta$.
In the E-step, we calculate the expectation of the
complete-data log-likelihood under the distribution $P(\mathbf{z},
\mathbf{X} \vert \bolds\Theta^{(t)}, \mathbf{Y} )$.
The marginalization of $\mathbf{X}$, conditional on
$\mathbf{z}$ and other variables, can be accomplished efficiently
site-by-site using the peeling or pruning algorithm for the
phylogenetic tree (\citet{refFelsenstein81JMolEvol}). The
marginalization of $\mathbf{z}$ can be done efficiently by the
forward--backward procedure for HMM (\citeauthor{reBaumWeiss70AMS},
\citeyear{reBaumWeiss70AMS};
\citeauthor{refRabine89ProcIEEE}, \citeyear{refRabine89ProcIEEE}).
For the M-step, we can use the
Broyden--Fletcher--Goldfarb--Shanno (BFGS) quasi-Newton algorithm.
After we obtain the MLE of $\bolds\Theta$, a
forward--backward dynamic programming method
(\citeauthor{refLiu01book}, \citeyear{refLiu01book}) can then be used
to compute the posterior
probability that a given hidden state is conserved, that is, $P(z_i
= c \vert \hat{\bolds\Theta}, \mathbf{Y} )$, which is the
desired conservation score.

As shown in the Phylo-HMM example, the phylogenetic tree model
is key to integrating multiple sequences for evolutionary
analysis. This model is also used for comparing protein or RNA
sequences. Due to its intuitive and efficient handling of the
missing evolutionary history, the EM algorithm has always been a
main approach for estimating parameters of the tree. For example,
\citet{refFelsenstein81JMolEvol} used the EM algorithm to
estimate the branch length $\bolds\beta$,
\citet{refBruno96MBE} and \citet{refHolmesRubin02JMB} used
the EM algorithm to estimate the residue usage $\pi$ and the
substitution rate matrix $Q$, \citet{refFriedmanPupko02JCB}
used an extension of the EM algorithm to estimate the phylogenetic
tree topology $\tau$, and \citet{refHolmes05Bioinfo} used the EM
algorithm for estimating insertion and deletion rates.
\citet{refYang97ComputApplBiosci} implemented some of the above
algorithms in the phylogenetic analysis software PAML. A
limitation of the Phylo-HMM model is the assumption of a good multiple
sequence alignment, which is often not available.

\section{SNP Haplotype Inference}\label{sec5}

\begin{figure}[b]
\vspace*{-2pt}
\includegraphics{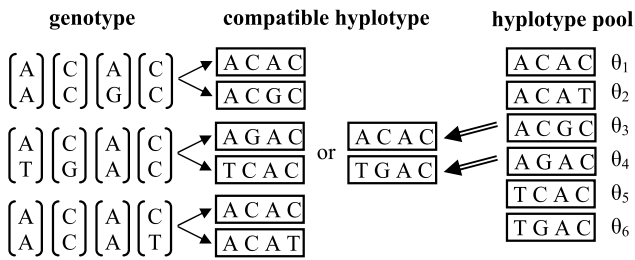}

\caption{Haplotype reconstruction. We observed the genotypes of
three individuals at 4 SNP loci. The 1st and 3rd individuals each have
a unique haplotype phase, whereas the 2nd individual has two
compatible haplotype phases. We pool all possible haplotypes
together and associated with them a haplotype frequency vector
$(\theta_1,\ldots ,\theta_6)$. Each individual's two haplotypes are
then assumed to be random draws (with replacement) from this pool of
weighted haplotypes.}
\label{fig:haplotyping} 
\end{figure}

A Single Nucleotide Polymorphism (SNP) is a DNA sequence variation
in which a single base is altered that occurs in at least 1\% of the
population. For example,
the DNA fragments CCTG\textit{\textbf{A}}GGAG and
CCTG\textit{\textbf{T}}GGAG from two homologous
chromosomes (the paired chromosomes of the same
individual, one from each parent) differ at a single locus. This
example is
actually a real SNP in the human $\beta$-globin gene, and it is
associated with the sickle-cell disease. The different forms (A
and T in this example) of a SNP are called alleles. Most SNPs have
only two alleles in the population. Diploid organisms, such as
humans, have two homologous copies of each chromosome. Thus, the
genotype (i.e., the specific allelic makeup) of an individual may
be AA, TT or AT in this example. A phenotype is a morphological
feature of the organism controlled or affected by a genotype.
Different genotypes may produce the same phenotype. In
this example, individuals with genotype TT have a very high risk
of the sickle-cell disease. A haplotype is a combination of
alleles at multiple SNP loci that are transmitted together on the
same chromosome. In other words, haplotypes are sets of \textit{phased}
genotypes. An example is given in Figure~\ref{fig:haplotyping}, which
shows the genotypes of three individuals at four
SNP loci. For the first individual, the arrangement of its alleles
on two chromosomes must be ACAC and ACGC, which are the haplotypes
compatible with its observed genotype data.

One of the main tasks of genetic studies is to locate genetic
variants (mainly SNPs) that are associated with inheritable
diseases. If we know the haplotypes of all related individuals, it
will be easier to rebuild the evolutionary history and locate the
disease mutations. Unfortunately, the phase information needed to
build haplotypes from genotype information is usually unavailable
because laboratory haplotyping methods, unlike genotyping technologies,
are expensive and low-throughput.

The use of the EM algorithm has a long history in population
genetics, some of which predates
\citet{refDempsterRubin77JRSSB}. For example,
\citet{refCeppelliniSmith55AHG} invented an EM algorithm to
estimate allele frequencies when there is no one-to-one
correspondence between phenotype and genotype;
\citet{refSmith57AHG} used an EM algorithm\vadjust{\goodbreak} to estimate the
recombination frequency; and \citet{refOtt79AJHG} used an EM
algorithm to study genotype-phenotype relationships from pedigree
data. \citet{refWeeksLANGE89MathMedBiol} reformulated these
earlier applications in the modern EM framework of
\citet{refDempsterRubin77JRSSB}. Most early works were single-SNP
Association studies.
\citet{refThompson84MMB} and \citet{refLanderGreen87PNAS}
designed EM algorithms for joint linkage analysis of three or more
SNPs. With the accumulation of SNP data, more and more researchers
have come to realize the importance of haplotype analysis
(\citeauthor{refLiuRisch01GenomeRes}, \citeyear{refLiuRisch01GenomeRes}).
Haplotype reconstruction
based on genotype data has therefore become a very important
intermediate step in disease association studies.

The haplotype reconstruction problem is illustra\-ted in
Figure~\ref{fig:haplotyping}. Suppose we observed the genotype data
$\mathbf{Y}=(Y_1,\ldots,Y_n)$ for $n$ individuals,
and we wish to predict the corresponding haplotypes
$\bolds\Gamma=(\Gamma_1,\ldots,\break\Gamma_n)$, where $\Gamma_i =
(\Gamma^+_i,\Gamma^-_i)$ is the haplotype pair of the
\textit{i}th individual. The haplotype pair $\Gamma_i$ is said to
be compatible with the genotype $Y_i$, which is expressed as
$\Gamma^+_i \oplus\Gamma^-_i = Y_i$, if the genotype $Y_i$ can
be generated from the haplotype pair. Let $\mathbf{H} =(H_1,
\dots, H_m)$ be the pool of all distinct haplotypes and let
$\bolds\Theta= (\theta_1, \ldots, \theta_m)$ be the
corresponding frequencies in the population.

The first simple model considered in the literature assumes that
each individual's genotype vector is generated by two haplotypes
from the pool chosen independently with probability vector
$\bolds{\Theta}$. This is a very good model if the region spanned by the
markers in consideration is sufficiently short that no recombination
has occurred, and if mating in the population is random.
Under this model, we have
%
\[
P(\mathbf{Y} \vert \bolds\Theta) = \prod^n_{i=1} \biggl(
\sum_{(j,k):H_j \oplus H_k = Y_i} \theta_j \theta_k \biggr).
\]
If $\bolds\Gamma$ is known, we can directly write down
the MLE of $\bolds\Theta$ as $\theta_j = \frac{n_j}{2n}$,
where the sufficient statistic $n_j$ is the number of occurrences
of haplotype $H_j$ in $\bolds\Gamma$. Therefore, in the EM
framework, we simply replace $n_j$ by its expected value over the
distribution of $\bolds\Gamma$ when $\bolds\Gamma$ is
unobserved. More specifically, the EM algorithm is a simple
iteration of
\[
\theta^{(t+1)}_j = \frac{E_{\bolds\Gamma\mid\bolds\Theta^{(t)},
\mathbf{Y}} (n_j)}{2n},
\]
where $\bolds\Theta^{(t)}$ is the current estimate of the
haplotype frequencies, and $n_j$ is the count of haplotypes $H_j$
that exist in $\mathbf{Y}$.

The use of the EM algorithm for haplotype analysis has been
coupled with the large-scale generation of SNP data. Early
attempts include \citet{refExcoffierSlatkin95MBE},
\citet{refLongUrbanek95AJHG},
\citet{reHawleyKidd95JHered} and
\citet{refChianoClayton98AHG}. One problem of these
traditional EM approaches is that the computational complexity of
the E-step grows exponentially as the number of SNPs
in the haplotype increases. \citet{refQinLiu02AJHG} incorporated
a ``partition--ligation'' strategy into the EM algorithm in an effort
to surpass this limitation. \citet{refLuLiu03GenomeRes} used
the EM for haplotype analysis in the scenario of case-control
studies. \citet{refKangLiu04AJHG} extended the traditional EM
haplotype inference algorithm by incorporating genotype
uncertainty. \citet{refNiu04GenetEpidem} gave a review of
general algorithms for haplotype reconstruction.

\section{Finite Mixture Clustering for Microarray
Data}\label{sec6}

In cluster analysis one seeks to partition observed data into
groups such that coherence within each group and separation
between groups are maximized jointly. Although this goal is
subjectively defined
(depending on how one defines ``coherence'' and ``separation''),
clustering can serve as an initial explorato\-ry analysis for
high-dimensional data. One example in computational biology is
microarray data analysis.
Microarrays are used to measure the mRNA expression levels of
thousands of genes at the same time. Microarray data are usually
displayed as a
matrix $\mathbf{Y}$. The rows of $\mathbf{Y}$ represent the
genes in a study and the columns are arrays obtained in different
experiment conditions, in different stages of a biological
system or from different biological samples. Cluster analysis of
microarray data has been a hot research field because
groups of genes that share similar expression patterns
(clustering the rows of $\mathbf{Y}$) are often involved in the
same or related biological functions, and groups of samples having
a similar gene expression profile
(clustering the columns of $\mathbf{Y}$) are often indicative of the
relatedness of these samples (e.g., the same cancer type).

Finite mixture models have long been used in cluster analysis
(see \citeauthor{refFraleyRaftery02JASA}, \citeyear{refFraleyRaftery02JASA}
for a review). The
observations are assumed to be generated from a finite mixture of
distributions. The likelihood of a mixture model
with $K$ components can be written as
\[
P(\mathbf{Y} | \bolds\theta_1,\ldots,\bolds\theta_K;
\tau_1,\ldots,\tau_K) = \prod_{i=1}^n\sum_{k=1}^K \tau_k
f_k(\mathbf{Y}_i|\bolds\theta_k),
\]
where $f_k$ is the density function of the $k$th component in the
mixture, $\bolds\theta_k$ are the corresponding parameters,
and $\tau_k$ is the probability that an observed datum is generated
from this component model ($\tau_k \ge0, \sum_k \tau_k=1$). One
of the most commonly used finite mixture models is the Gaussian
mixture model, in which $\bolds\theta_k$ is composed of mean
$\bolds\mu_k$ and covariance matrix $\bolds\Sigma_k$.
Outliers can be accommodated by a special
component in the mixture that allows for a larger variance or
extreme values.

A standard way to simplify the statistical computation with
mixture models is to introduce a variable indicating which
component an observation $\mathbf{Y}_i$ was generated from. Thus, the
``complete data'' can be expressed as $\mathbf{X}_i =
(\mathbf{Y}_i, \bolds\Gamma_i)$, where $\bolds\Gamma_i = (\gamma_{i1},
\ldots ,\gamma_{iK})$, and \mbox{$\gamma_{ik}=1$} if
$\mathbf{Y}_i$ is generated by the $k$th component and
$\gamma_{ik}=0$ otherwise.
The complete-data log-likelihood function is
\begin{eqnarray*}
&&\log P(\mathbf{Y}, \bolds\Gamma|\bolds\theta_1,\ldots ,\bolds\theta_K;
\tau_1,\ldots ,\tau_K)\\
&&\quad  =
\sum_{i=1}^n \sum_{k=1}^K \gamma_{ik}\log[\tau_k f_k(\mathbf{Y}_i|\bolds\theta_i)].
\end{eqnarray*}

Since the complete-data log-likelihood function is linear in
the $\gamma_{jk}$'s, in the E-step we only need to compute
%
\[
\hat{\gamma}_{ik} \equiv E\bigl(\gamma_{ik}|\bolds\Theta^{(t)},\mathbf{Y}\bigr) =
\frac{\tau_k^{(t)} f_k(\mathbf{Y}_i|\bolds\theta_k^{(t)})}
{\sum_{j=1}^K \tau_j^{(t)}
f_j(\mathbf{Y}_i|\bolds\theta_j^{(t)})}.
\]
The $Q$-function can be calculated as
%
\begin{equation}\label{mixture-q}
Q\bigl(\bolds\Theta|\bolds\Theta^{(t)}\bigr) = \sum_{i=1}^n
\sum_{k=1}^K \hat{\gamma}_{ik} \log[\tau_k f_k(\mathbf{Y}_i|\bolds\theta_i)].
\end{equation}
The M-step updates the component probability $\tau_k$ as
\[
\tau_k^{(t+1)} = \frac{1}{n} \sum_{i=1}^n \hat{\gamma}_{ik},
\]
and the updating of $\bolds\theta_k$ would depend on the
density function. In mixture Gaussian models, the $Q$-function is
quadratic in the mean vector and can be maximized to achieve the
M-step.

\citet{refYeungRuzzo01Bioinfo} are among the pioneers who
applied the model-based clustering method in microarray data analysis.
They adopted the Gaussian mixture model framework and represented
the covariance matrix in terms of its eigenvalue decomposition
\[
\bolds{\Sigma}_k = \lambda_k D_k A_k D_k^T.
\]
In this way, the orientation, shape and volume of the multivariate
normal distribution for each cluster can be modeled separately by
eigenvector matrix $D_k$, eigenvalue matrix $A_k$ and scalar
$\lambda_k$, respectively. Simplified models are straightforward
under this general model setting, such as setting $\lambda_k$,
$D_k$ or $A_k$ to be identical for all clusters or restricting the
covariance matrices to take some special forms (e.g., $ \bolds{\Sigma}_k =
\lambda_k I$). Yeung and colleagues used the EM algorithm to estimate
the model parameters. To improve convergence, the EM algorithm can be
initialized with a model-based hierarchical clustering step
(\citeauthor{refDasguptaRaftery98JASA}, \citeyear{refDasguptaRaftery98JASA}).

When $\mathbf{Y}_i$ has some dimensions that are highly correlated,
it can be helpful to
project the data onto a lower-dimensional subspace. For example,
\citet{refMcLachlanPeel02Bioinfo} attempted to cluster tissue
samples instead of genes. Each tissue sample is represented as a vector of
length equal to the number of genes, which can be up to several
thousand. Factor analysis (\citeauthor{refGhahramaniHinton97TechReport},
\citeyear{refGhahramaniHinton97TechReport})
can be used to reduce the
dimensionality, and can be seen as a Gaussian model with a
special constraint on the covariance matrix. In their study, McLachlan,
Bean and Peel
used a mixture of factor analyzers, equivalent to a mixture
Gaussian model, but with fewer free parameters to
estimate because of the constraints. A variant of the EM
algorithm, the Alternating Expectation--Conditional Maximization
(AECM) algorithm (\citeauthor{refMengvanDyk97JRSSB},
\citeyear{refMengvanDyk97JRSSB}), was applied to
fit this mixture model.

Many microarray data sets are composed of several arrays in a
series of time points so as to study biological system dynamics
and regulatory networks (e.g., cell cycle studies). It is advantageous to
model the gene expression profile by taking into account the
smoothness of these time series. \citet{refJiLi04Bioinfo}
clustered the time course microarray data using a mixture of HMMs.
\citet{refBar-JosephSimon02RECOMB} and
\citet{refLuanLi03Bioinfo} implemented mixture models with
spline components. The time-course expression data were treated as
samples from a continuous smooth process. The coefficients of the
spline bases can be either fixed effect, random effect or a mixture
effect to accommodate different modeling
needs. \citet{refMaLiu06NAR} improved upon these methods by adding
a gene-specific effect into the model:
\[
y_{ij} = \mu_k(t_{ij}) + b_i + \varepsilon_{ij},
\]
where $\mu_k(t)$ is the mean expression of cluster $k$ at time~$t$,
composed of smoothing spline components; $b_i \sim N(0,
\sigma_{bk}^2)$ explains the gene specific deviation from the
cluster mean; and $\varepsilon_{ij} \sim N(0,\sigma^2)$ is the
measurement error. The $Q$-function in this case is a weighted
version of the penalized log-likelihood:
\begin{eqnarray}\label{spline}
&&-\sum_{k=1}^K \Biggl\{ \sum_{i=1}^n \hat{\gamma}_{ik} \Biggl(
\sum_{j=1}^T \frac{(y_{ij}-\mu_k(t_{ij}) - b_i)^2}{2\sigma^2} +
\frac{b_i^2}{2\sigma_{bk}^2} \Biggr)\nonumber\\[-8pt]\\[-8pt]
&&\hspace*{130pt}{} - \lambda_k T \int
[\mu_k''(t)]^2 \,dt \Biggr\},\nonumber
\end{eqnarray}
where the integral is the smoothness penalty term. A
generalized cross-validation method was applied to choose the
values for $\sigma_{bk}^2$ and $\lambda_k$.

An interesting variation on the EM algorithm, the
rejection-controlled EM (RCEM), was introduced in
\citet{refMaLiu06NAR} to reduce the computational complexity
of the EM algorithm for mixture models. In all mixture models, the
E-step computes the membership probabilities (weights) for each
gene to belong to each cluster, and the M-step maximizes a
weighted sum function as in \citet{refLuanLi03Bioinfo}. To
reduce the computational burden of the M-step, we can ``throw
away'' some terms with very small weights in an unbiased weight
using the rejection control method (\citeauthor{refLiuWong98JASA},
\citeyear{refLiuWong98JASA}).
More precisely, a
threshold $c$ (e.g., $c=0.05$) is chosen.
Then, the new weights are computed as
%
\[
\tilde{\gamma}_{ik} =
\cases{
\max\{ \hat{\gamma}_{ik}, c\}, & with probability $\min\{1,\hat{\gamma}_{ik}/c\}$,
\vspace*{2pt}
\cr
0, & otherwise.}
\]
The new weight $\tilde{\gamma}_{ik}$ then replaces
the old weight $\hat{\gamma}_{ik}$ in
the $Q$-function calculation in (\ref{mixture-q}) in general, and in
(\ref{spline}) more specifically. For cluster $k$, genes
with a membership probability higher than $c$ are not affected,
while the membership probabilities of other genes will be set to
$c$ or 0, with probabilities $\hat{\gamma}_{ik}/c$ and
$1-\hat{\gamma}_{ik}/c$, respectively. By giving a zero weight to
many genes with low $\hat{\gamma}_{ik}/c$, the number of terms to
be summed in the $Q$-function is greatly reduced.

In many ways finite mixture models are similar to the K-means algorithm,
and they may produce very similar clustering results. However, finite
mixture models are more flexible in the sense that the inferred
clusters do not
necessarily have a sphere shape, and the shapes of the clusters can be learned
from the data. Researchers such as \citet{refSureshValarmathie09ICIME}
tried to combine the two ways of thinking to make better clustering
algorithms.

For cluster analysis, one intriguing question is how to set the total
number of clusters. Bayesian information criterion (BIC) is often used
to determine
the number of clusters (\citet{refYeungRuzzo01Bioinfo};
\citet{refFraleyRaftery02JASA};
\citet{refMaLiu06NAR}). A random
subsampling approach is suggested by
\citet{refDudoitSpeed02JASA} for the same purpose. When
external information of genes or samples is available, cross-validation
can be used to determine the number of clusters.

\section{Trends Toward Integration}\label{sec7}

Biological systems are generally too complex to be fully
characterized by a snapshot from a single viewpoint. Modern
high-throughput experimental techniques have been used to collect
massive amounts of data to interrogate biological systems from
various angles and under diverse conditions. For instance, biologists
have collected many types of genomic data,
including microarray gene expression data, genomic sequence data,
ChIP--chip binding data and protein--protein interaction data.
Coupled with this trend, there is a growing interest in
computational methods for integrating multiple sour\-ces of
information in an effort to gain a deeper understanding of the
biological systems and to overcome the limitations of divided
approaches. For
example, the Phylo-HMM in Section~\ref{sec4} takes
as input an alignment of multiple sequences, which, as shown in
Section~\ref{sec3}, is a hard problem by itself. On the
other hand, the construction of the alignment can be improved a lot if
we know the underlying phylogeny. It is therefore preferable to infer
the multiple alignment and the phylogenetic tree jointly
(\citeauthor{refLunterHein05BMCBioinfo}, \citeyear{refLunterHein05BMCBioinfo}).

Hierarchical modeling is a principled way of integrating multiple
data sets or multiple analysis steps. Because of the complexity of the
problems, the inclusion of nuisance parameters or missing data at
some level of the hierarchical models is usually either structurally
inevitable or conceptually preferable. The EM algorithm and Markov
chain Monte Carlo algorithms are often the methods of choice for
these models due to their close connection with the underlying
statistical model and the missing data structure.

For example, EM algorithms have been used to combine motif discovery with
evolutionary information. The underlying logic is that the motif
sites such as TFBSs evolved slower than the
surrounding genomic sequences (the background) because of functional
constraints and natural
selection. \citet{refMosesEisen04PSB} developed EMnEM
(Expecta\-tion--Maximization on Evolutionary Mixtures), which is a
generalization of the mixture model formulation for motif
discovery (\citeauthor{refBaileyElkan94ISMB}, \citeyear{refBaileyElkan94ISMB}).
More precisely, they
treat an alignment of multiple orthologous sequences as a series
of alignments of length $w$, each of which is a sample from the
mixture of a motif model and a background model. All observed sequences
are assumed to evolve from a common ancestor sequence according to
an evolutionary process parameterized by a
Jukes--Cantor substitution matrix. PhyME
(\citeauthor{refSinhaTompa04BMCBioinfo}, \citeyear{refSinhaTompa04BMCBioinfo})
is another EM approach for
motif discovery in orthologous sequences. Instead of modeling the
common ancestor, they modeled one designated ``reference species''
using a two-state HMM (motif state or background state). Only the
well-aligned part of the reference sequence was assumed to share
a common evolutionary origin with other species. PhyME assumes a
symmetric star topology instead of a binary phylogenetic tree for
the evolutionary process. OrthoMEME
(\citeauthor{refPrakashTompa04PSB}, \citeyear{refPrakashTompa04PSB}) deals
with pairs of orthologous
sequences and is a natural extension of the EM algorithm of
\citet{refLawrenceReilly90Proteins} described in
Section~\ref{sec2}.

Steps have also been taken to incorporate micro\-array gene
expression data into motif discovery\break
(\citet{refBussemakerSiggia01NatureGenet};
\citet{refConlonLiu03PNAS}).
\citet{refKundajeLeslie05IEEECompBio} used a graphical model
and the EM algorithm to combine DNA sequence data with time-series
expression data for gene clustering. Its basic logic
is that co-regulated genes should show both similar TFBS
occurrence in their upstream sequences and similar gene-expression
time-series curves. The graphical model assumes that the TFBS
occurrence and gene-expression are independent, conditional on the
co-regulation cluster assignment. Based on predicted TFBSs in
promoter regions and cell-cycle time-series gene-expression data
on budding yeast, this algorithm infers model parameters by
integrating out the latent variables for cluster
assignment. In a similar setting,
\citet{refChenBlanchette07BMCBioinfo} used a Bayesian network
and an EM-like algorithm to integrate TFBS information, TF
expression data and target gene expression data for identifying
the combination of motifs that are responsible for
tissue-specific expression. The relationships among different data are
modeled by the connections of different nodes in the Bayesian network.
\citet{refWangLi05PNAS} used a mixture model to describe the
joint probability of TFBS and target gene expression data. Using
the EM algorithm, they provide a refined representation of the
TFBS and calculate the probability that each gene is a true
target.

As we show in this review, the EM algorithm has enjoyed many
applications in computational biology. This is partly driven by
the need for complex statistical models to describe biological
knowledge and data. The missing data formulation of the EM
algorithm addresses many computational biology problems naturally.
The efficiency of a specific EM algorithm depends on how
efficiently we can integrate out unobserved variables (missing
data/nuisance parameters) in the E-step and how complex the
optimization problem is in the M-step. Special dependence
structures can often be imposed on the unobserved variables to
greatly ease the computational burden of the E-step. For example,
the computation is simple if latent variables are independent in
the conditional posterior distribution, such as in the mixture
motif example in Section~\ref{sec2} and the haplotype
example in Section~\ref{sec5}. Efficient exact
calculation may also be available for structured latent variables,
such as the forward--backward procedure for HMMs
(\citeauthor{reBaumWeiss70AMS}, \citeyear{reBaumWeiss70AMS}),
the pruning algorithm for
phylogenetic trees (\citeauthor{refFelsenstein81JMolEvol},
\citeyear{refFelsenstein81JMolEvol}) and the
inside--outside algorithm for the probabilistic context-free
grammar in predicting RNA secondary structures
(\citeauthor{refEddyDurbin94NAR}, \citeyear{refEddyDurbin94NAR}).
As one of the drawbacks of the EM
algorithm, the M-step can sometimes be too complicated to compute
directly, such as in the Phylo-HMM example in
Section~\ref{sec4} and the smoothing spline mixture
model in Section~\ref{sec6}, in which cases
innovative numerical tricks are called for.

\section*{Acknowledgments}
We thank Paul T. Edlefsen for helpful discussions about the profile
hidden Markov model, as well as to Yves Chretien for polishing the language.
This research is supported in part by the NIH
Grant R01-HG02518-02 and the NSF Grant DMS-07-06989.
The first two authors should be regarded as joint first authors.

\end{document}